\input psfig.sty
\documentclass{kapproc} % Computer Modern font calls
\kluwerbib
\begin{document}
\articletitle{FIR and Radio emission in \\
star forming galaxies}
%\articlesubtitle{star forming galaxies}
\author{Alessandro Bressan$^1$,
Gian Luigi Granato$^1$, Laura Silva$^2$}
\affil{$^1$Osservatorio Astronomico di Padova, Padova,IT}
\affil{$^2$Osservatorio Astronomico di Trieste, Trieste,IT}
\begin{abstract}
We examine the tight correlation observed between the far infrared
(FIR) and radio emission in normal and star burst galaxies. We show that
significant deviations from the average relation are to be expected in
young star burst or post star burst galaxies, due to the different
fading times of FIR and Radio power. 
\end{abstract}

\section{The Q-slope diagram}
\label{sec:qslo}
Observations indicate that there is a tight correlation
between the FIR and Radio emission among star forming galaxies
(Sanders and Mirabel, 1996). At 1.4GHz, 
$ Q =\log \frac{FIR/(3.75\times 10^{12}erg~s^{-1})}{L\nu
(1.4GHz)/(erg~s^{-1}Hz^{-1})}\simeq 2.3\pm 0.2$.
However,  significant deviations from this relation, though with a larger
scatter, are observed either in ultra luminous infrared
galaxies (ULIRGS), where Q may reach values as high as 3 (Condon et al, 1991), or in
the central regions of rich clusters where there seem to be a significant
excess of star forming galaxies with Q below 2 (Miller and Owen, 2001).
To cast light on these problems, we have analysed the
evolution of FIR and Radio emission of a starburst model, consisting
of a superposition of
an exponentially declining star-burst episode to a quiescent disk galaxy. 
The burst is described by
its strength, its age $T_b$ and the e-folding 
time $\tau_b$.
Thermal radio emission is related to the ionizing photon flux from young stars,
while non thermal emission, 
-$L_{NT}(\nu)= L_{1.4}^{SNR}(\frac{\nu}{1.4})^{-0.5} +
K \times \nu _{SN} (\frac{\nu}{1.4})^{-\alpha}$-
consisting of the contribution from supernova remnants
and an unidentified mechanism, 
is normalized to the ratio between
NT radio emission and supernova rate, observed  in the Galaxy (Condon, 1992). 
By adopting $\nu_{SN}\simeq~0.018$ (Cappellaro, private communication)
and 
$L_{0.4GHz}~\simeq 6.1\times~10^{28} erg/s/Hz$ 
Berkhuijsen (1984), we obtain $K\simeq~1.04 \ 10^{30}$erg/s/Hz/Yr$^{-1}$.
In order to reproduce the observed average slope of the NT radio emission in normal
spiral galaxies, $\alpha\simeq$~0.9.
The infrared emission depends on the time (t$_{FIR}$) by which
young populations get rid of their parental molecular clouds
(eg. Silva et al 1998). We have found that with an escape time
of t$_{FIR}\simeq$3Myr and a visual extinction of A$_V$=1mag, we reproduce well the
FIR/Radio correlation observed in normal star forming galaxies -constant SFR 
and Salpeter IMF between 0.15 and 120M$_\odot$, over the last few 100 Myr- namely
Q$_{1.4GHz}$=2.3.
The adopted values are consistent with independent results 
of recent infrared modelling of spiral galaxies.
With this model we also obtain 
$Sfr\simeq 0.57 \ 10^{-28}F_{1.4GHz}(erg/s/Hz)$,
in good agreement with the one obtained by Carilli (2001). 
\begin{figure}[]
\psfig{file=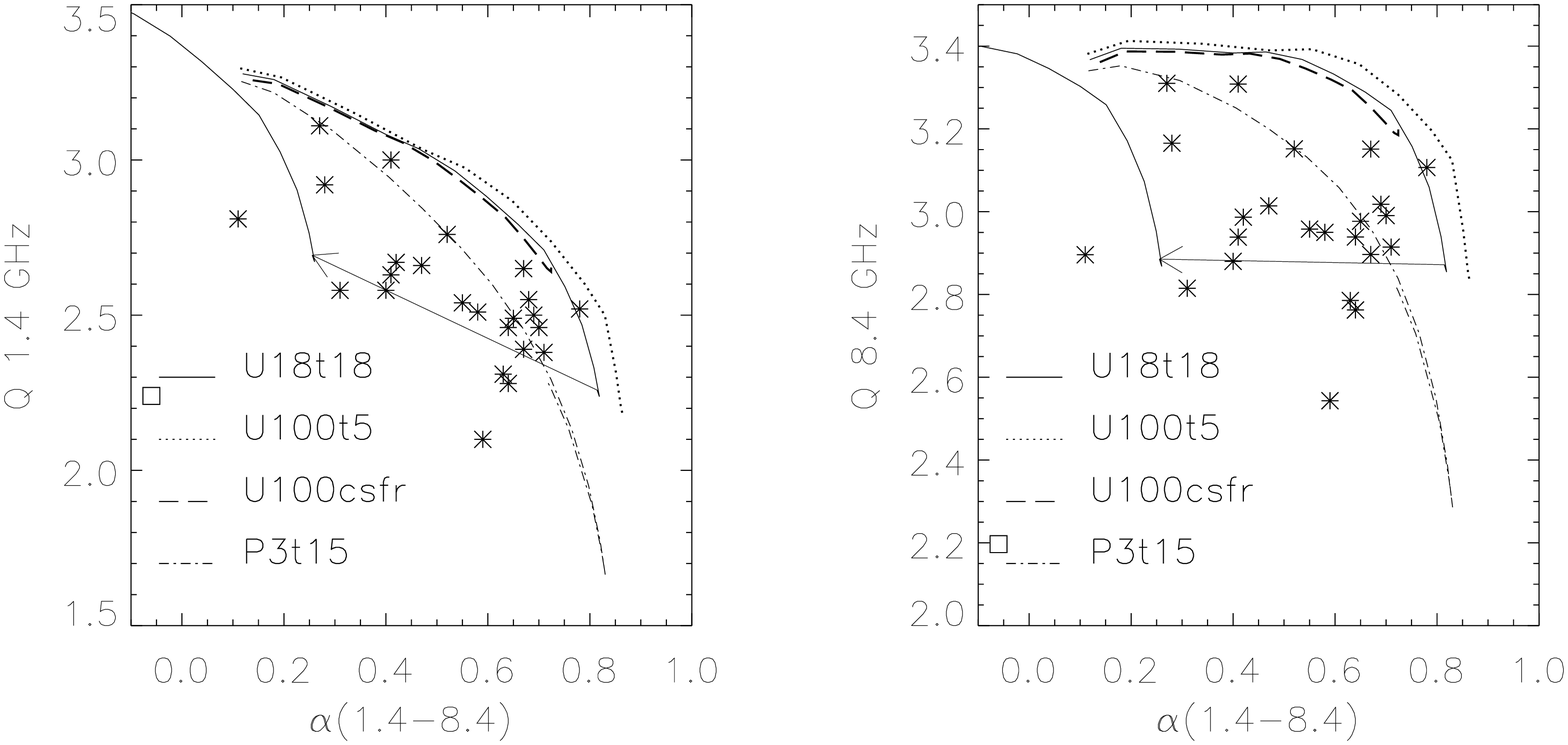,height=5.1cm,width=11.5cm}
\caption{Predicted evolution of Q vs radio slope ($\alpha=-\frac{dlogL_\nu}{dlog\nu}$) between 1.4 and 8.4 GHz, 
compared with observed values (stars) in compact ULIRGS. 
The burst age increases from left, log t(yr)=6.5, to right, log t(yr)$\leq$8.
U18 and U100 refer to models with t$_{FIR}$=18 and 100Myr while P3 to
post-starburst galaxies with t$_{FIR}$=3Myr. 
The last number is the e-folding time (Myr) of the SFR and "csfr" indicates constant SFR.
The arrow shows how free-free absorption, with $\tau_{1.4GHz}=1$, affects model U18t18.}
\label{allslo}
\end{figure}
In the analysis of a sample of compact ULIRGS (Condon et al, 1991), we adopt
larger obscuration times and a visual extinction of A$_V$=10mag, as
implied by optical and UV properties of observed SEDs of a few ULIRGS.
With the help
of a new diagnostic diagram, which contrasts the Q ratio with the radio slope 
(Fig.\ref{allslo}), we may draw the following conclusions. 
Our star burst models may account for the infrared and radio emission
of the observed ULIRGS. The mass involved in the burst, for the sample analysed is 
typically between 10$^9$M$_\odot$ and 10$^{10}$M$_\odot$.
The range of Q$_{1.4GHz}$ 
is due to a combined effect
of free-free absorption (see also Condon et al. 1991)
and real star burst evolution.
Free-free absorption induces an evident trend in the Q vs radio-slope diagram,
with the higher Q being accompanied by a shallower slope. 
Estimated optical depths for free-free absorption at 1.4GHz are
between 0.2 and 1. The derived emission measures are consistent with
the typical small sizes deduced from high resolution radio and mid
infrared images (Condon et al. 1991, Soifer et al. 2000) and the typical
ionized gas masses involved in the model star burst episode.
At 8.4GHz free-free absorption becomes negligible
($\tau_\nu\propto{\nu^{-2.1}}$) and the previous trend disappears.
The value of Q$_{8.4GHz}$ is a measure of the age of the
star burst. Very young star bursts display an excess of FIR emission
relative to the radio emission  because the latter is initially
contributed mainly by the free-free process. As the star burst ages,
the non-thermal contribution increases and becomes the dominant source.
The relative contribution of radio emission increases and the ratio
Q$_{8.4GHz}$ decreases.
We suggest that a similar diagram between
8.4GHz and a higher frequency would better highlight 
the evolutionary status of  compact ULIRGS because the slope, 
unaffected by free-free absorption, would provide 
an independent estimate of the age.
Models with constant star formation during the burst,
always have a value of Q$_{8.4GHz}$ which remains too high at later times.
As the only viable alternative, the star formation rate {\sl must}
decline strongly with time.
Models with t$_{FIR}$=18Myr and $\tau_{burst}$=18Myr or
t$_{FIR}$=100Myr and $\tau_{burst}$=5Myr reproduce the observed data. 
Apparently, the star formation in ULIRGS is strongly peaked and the
burst is younger than about 60 Myr.
The determination of the star formation rate from common
{\sl average} relations is obviously misleading in this evolutionary scenario. 
The open square in Fig.\ref{allslo} refers to 
the Seyfert 1 galaxy  MRK 231.
By looking to the offset
with respect to the other objects (or the obscured models with the lowest
allowed Q) we estimate that the AGN is contributing to 
the 75\% at least, of its total radio emission.
In Fig.\ref{allslo} we also show a post starburst model
with the same obscuration of normal spirals, but a peaked
SFR. We thus guess that a low Q in star forming cluster galaxies is a signature
of a dying star formation, recently ($\leq$150Myr ago) 
interrupted by environmental effects.

\smallskip
\noindent We acknowledge discussions with C. Gruppioni, B. Poggianti,  A. Franceschini
and support from the TMR grant ERBFMRX-CT96-0086.     
%\end{acknowledgments}

\end{document}